\documentclass[prl,noshowpacs,english,superscriptaddress,twocolumn]{revtex4}
\usepackage{amsmath,amssymb,amsfonts,mathrsfs}
\usepackage{amsthm}
\usepackage{CJK}
\usepackage{bm}
\usepackage{babel}
\usepackage{graphicx}

\begin{document}

\title{Ferromagnetic Type-II Weyl Semimetal in Pyrite Chromium Dioxide}

\author{R. Wang}
\affiliation{Department of Physics, South University of Science and Technology of China,
Shenzhen 518055, P. R. China.}
\affiliation{Institute for Structure and Function $\&$
Department of physics, Chongqing University, Chongqing 400044, P. R. China.}
\author{Y. J. Jin}
\affiliation{Department of Physics, South University of Science and Technology of China,
Shenzhen 518055, P. R. China.}
\author{J. Z. Zhao}
\affiliation{Department of Physics, South University of Science and Technology of China, Shenzhen 518055, P. R. China.}
\affiliation{Dalian Institute of Chemical Physics,Chinese Academy of Sciences, 116023 Dalian, P. R. China}
\author{Z. J. Chen}
\affiliation{Department of Physics, South University of Science and Technology of China, Shenzhen 518055, P. R. China.}
\affiliation{Department of Physics, South China University of Technology, Guangzhou 510640, P. R. China}
\author{Y. J. Zhao}
\affiliation{Department of Physics, South China University of Technology, Guangzhou 510640, P. R. China}
\author{H. Xu}
\email[]{xuh@sustc.edu.cn}
\affiliation{Department of Physics, South University of Science and Technology of China, Shenzhen 518055, P. R. China.}

\begin{abstract}
Magnetic topological materials have recently drawn significant importance and interest, due to their topologically nontrivial electronic structure within spontaneous magnetic moments and band inversion.  Based on first-principles calculations, we propose that chromium dioxide, in its ferromagnetic pyrite structure, can realize one pair of type-II Weyl points between the $N$th and $(N+1)$th bands, where $N$ is the total number of valence electrons per unit cell. Other Weyl points between the $(N-1)$th and $N$th bands also appear close to the Fermi level due to the complex topological electronic band structure. The symmetry analysis elucidates that the Weyl points arise from a triply-degenerate point splitting due to the mirror reflection symmetry broken in the presence of spin-orbital coupling, which is equivalent to an applied magnetic field along the direction of magnetization. The Weyl points located on the magnetic axis are protected by the three-fold rotational symmetry. The corresponding Fermi arcs projected on both (001) and (110) surfaces are calculated as well and observed clearly. This finding opens a wide range of possible experimental realizations of type-II Weyl fermions in a system with time-reversal breaking.

\end{abstract}



\maketitle

 Topologically protected fermions, appearing in semimetallic or metallic systems with nontrivial topology of band structure, provide a realistic platform for the concepts of fundamental physics theory in condensed matter experiments. For instance, the recent discoveries of Weyl semimetals (WSMs) have been attracted considerable attention since they extend the topological classification of mater beyond the insulators and exhibit the exotic Fermi arc surface state \cite{Wan2011, Xu2011}. In these materials, the valence and conduction bands disperse linearly around special points in three-dimensional (3D) momentum space, called Weyl points (WPs), which construct the discrete point-like Fermi surface. The WP acts as a topological monopole which can be quantified by corresponding chiral charge through calculating the flux of Berry curvature \cite{Wan2011, Weng2015,Huang2015}. Due to the conservation of chirality, the WPs always appear in pairs of opposite chirality. The topological Fermi arcs are arising from the connection of two projections of the bulk WPs with opposite chiral charges in the surface Brillouin zone (BZ). WSMs and their surface states may lead to unusual spectroscopic and transport phenomena such as chiral anomaly, spin and anomalous Hall effects \cite{Xu2011,Adler1969, Shekhar2015, Sun2016}.

WPs are twofold degeneracy and only exist in condensed matter systems with breaking either time-reversal or spatial-inversion symmetry. Evidences for Weyl fermions and surface Fermi arc states with breaking spatial-inversion were reported in the non-centrosymmetric TaAs family \cite{Xu2015,Lv2015,Lv2015NP} and MoTe$_2$ \cite{Tamai2016}. WSMs with time-reversal-breaking were also been predicted to exist in several materials \cite{Wan2011, Xu2011,Wang2016prl1,Borisenko,Chinotti2016prb,Hirschberger2016nm}. Moreover, the existence of two distinct types of WSMs was recently proposed. The type-I WP is associated with a closed pointlike Fermi surface, while the type-II one arises at the boundary of electron and hole pockets \cite{Soluyanov2015,Autes2016,Wang2016prl2}. Unlike the low-energy excitations in type-I WSMs, the type-II Weyl fermions don't satisfy Lorentz invariance, leaving an open Fermi surface that results in anisotropic chiral anomaly \cite{Soluyanov2015}. To date several nonmagnetic Type-II WSMs have been predicted \cite{Autes2016,Wang2016prl2} and observed in MoTe$_2$ \cite{Tamai2016}, while antiferromagnetic (AFM) YbMnBi$_2$ are found to be the candidate of type-II WSM with lacking time-reversal symmetry \cite{Borisenko,Chinotti2016prb}. Type-II WSMs co-existing with the ferromagnetic (FM) order have not been reported so far. The novel properties of FM type-II WSMs can look forward to be of great use for spin manipulation and applications in spintronics and magnetic recording devices.

In this Letter, based on first-principles calculations, we propose the pyrite chromium dioxide (CrO$_2$) that can exhibit FM type-II WSM features up to room temperature. It is noteworthy
that CrO$_2$ is a very common material in practice. Rutile CrO2 is a well-studied half-metallic FM material with a high Curie temperature of about $\sim$390 K \cite{Schwarz1986,Katsnelson2008}. Crystalline CrO$_2$ hosts a number of pressure-induced structural phases \cite{Kim2012,Kuznetsov,Li2012}. The pyrite CrO$_2$ phase has been demonstrated to be stable FM half-metallic state occurring at a critical pressure of $\sim$45 GPa \cite{Li2012}. As a significant advantage, our calculations show that the magnetism in pyrite CrO$_2$ is "soft", indicating that an extra magnetic field can easily change the direction of magnetization. Some similar phenomena are observed in magnetic Heusler alloys \cite{Wang2016prl1,Hirschberger2016nm}. As a result, the number and position of WPs depend on the magnetic symmetry, and then the topological features in pyrite CrO$_2$ can be manipulated.  Our symmetry analysis elucidates that the WPs in pyrite CrO$_2$ arise from a triply-degenerate point splitting in the presence of spin-orbital coupling (SOC), which is equivalent to an applied magnetic field along the direction of magnetization \cite{Lv2017,Zhuprx2016,Changarxiv}. This finding would provide a realistic and promising platform for investigating FM Weyl physics, especially opening a pathway for studying the quantum anomalous Hall effect in FM WSMs in experiments.

\begin{figure}
	\centering
	\includegraphics[scale=0.25]{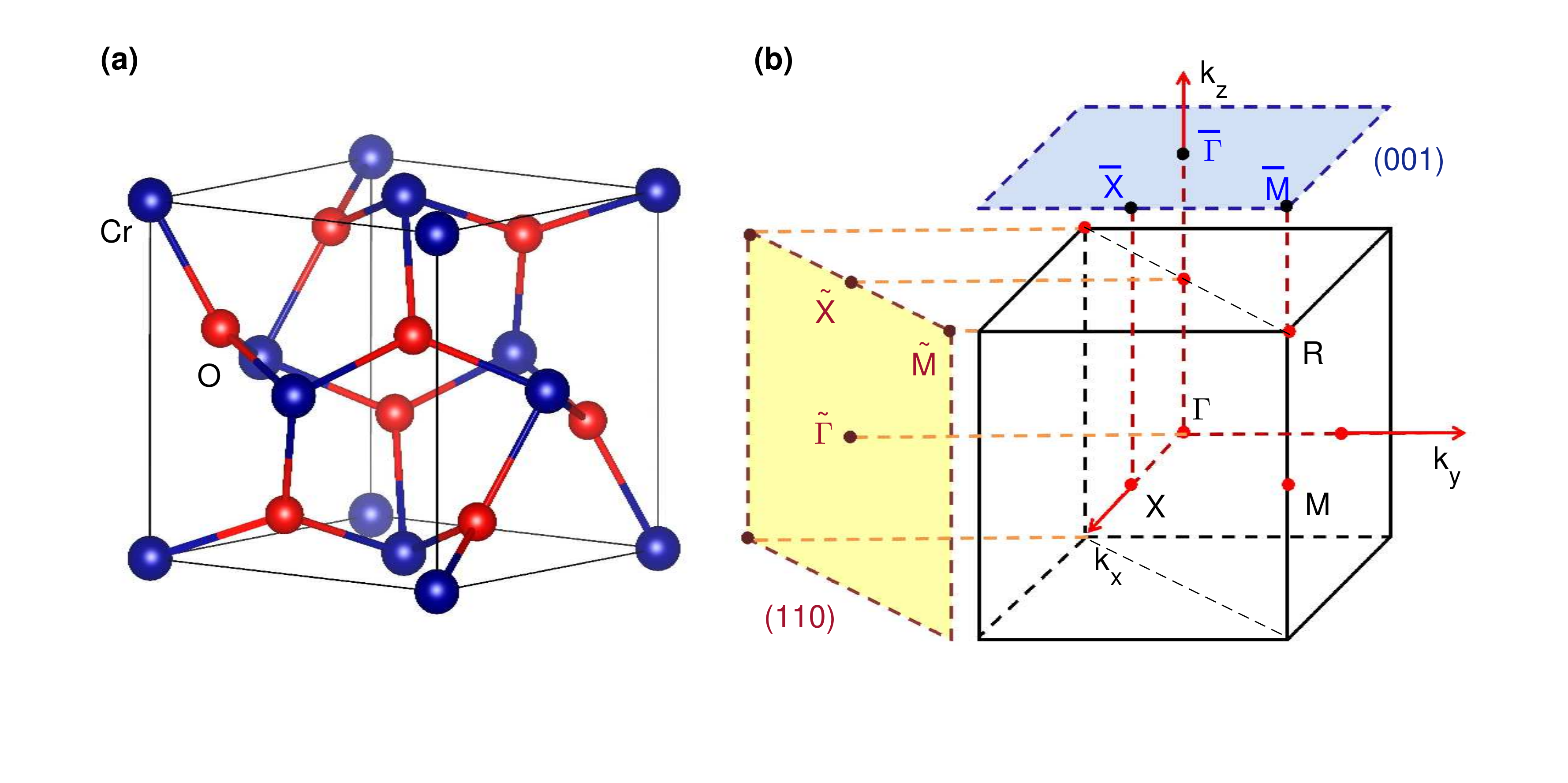}
	\caption{Crystal structure and Brillouin zone (BZ). (a) Crystal structure of pyrite CrO$_2$ with symmorphic space group $Pa\bar{3}$ (No. 205). Cr and O atoms are indicated by blue and red spheres, respectively. (b) The bulk BZ and the projected surface BZ for both (001) and (110) surfaces. The relevant high symmetry points are also indicated.
\label{figure1}}
\end{figure}

We perform the first-principles calculations using the Vienna \textit{ab initio} Simulation Package (VASP) \cite{Kresse2,Kressecom} based on density functional theory \cite{Hohenberg,Kohn}. The core-valence interactions are treated by the projector augmented wave (PAW) \cite{Blochl,Kresse4,Ceperley1980} pseudopotentials with $4p^6 4s^1 3d^5 $ and $2s^2 2p^4$ valence electron configurations for Cr and O, respectively. SOC effect is included in the pseudopotentials, and the exchange-correlation potential is chosen as generalized gradient approximation (GGA) with the Perdew-Burke-Ernzerhof (PBE) formalism \cite{Perdew1,Perdew2}. A plane-wave-basis set with kinetic-energy cutoff of 600 eV has been used. The full Brillouin zone(BZ) is sampled by $21\times21\times21$ Monkhorst-Pack grid in self-consistent calculations \cite{Monkhorst}.  Because of the strongly correlated effects of $3d$ electrons in Cr, we must consider the GGA+U calculations to describe the on-site Coulomb repulsion beyond the GGA calculations \cite{Liechtenstein1995,Korotin1998}. In this work, we find that topological features can be achieved with the range of U from 2.5 eV to 5.0 eV. As a representative, the value of U is chosen to be 3.5 eV to illustrate the band topology, which works well in fitting the half-metallic properties of rutile CrO$_2$ \cite{Katsnelson2008}. To calculate the surface states and Fermi arcs, the tight-binding Hamiltonian for Cr $3d$ and O $2p$ orbitals is constructed by projecting the Bloch states into maximally localized Wannier functions \cite{Marzari2012,Mostofi2008}.

\begin{figure}
	\centering
	\includegraphics[scale=0.35]{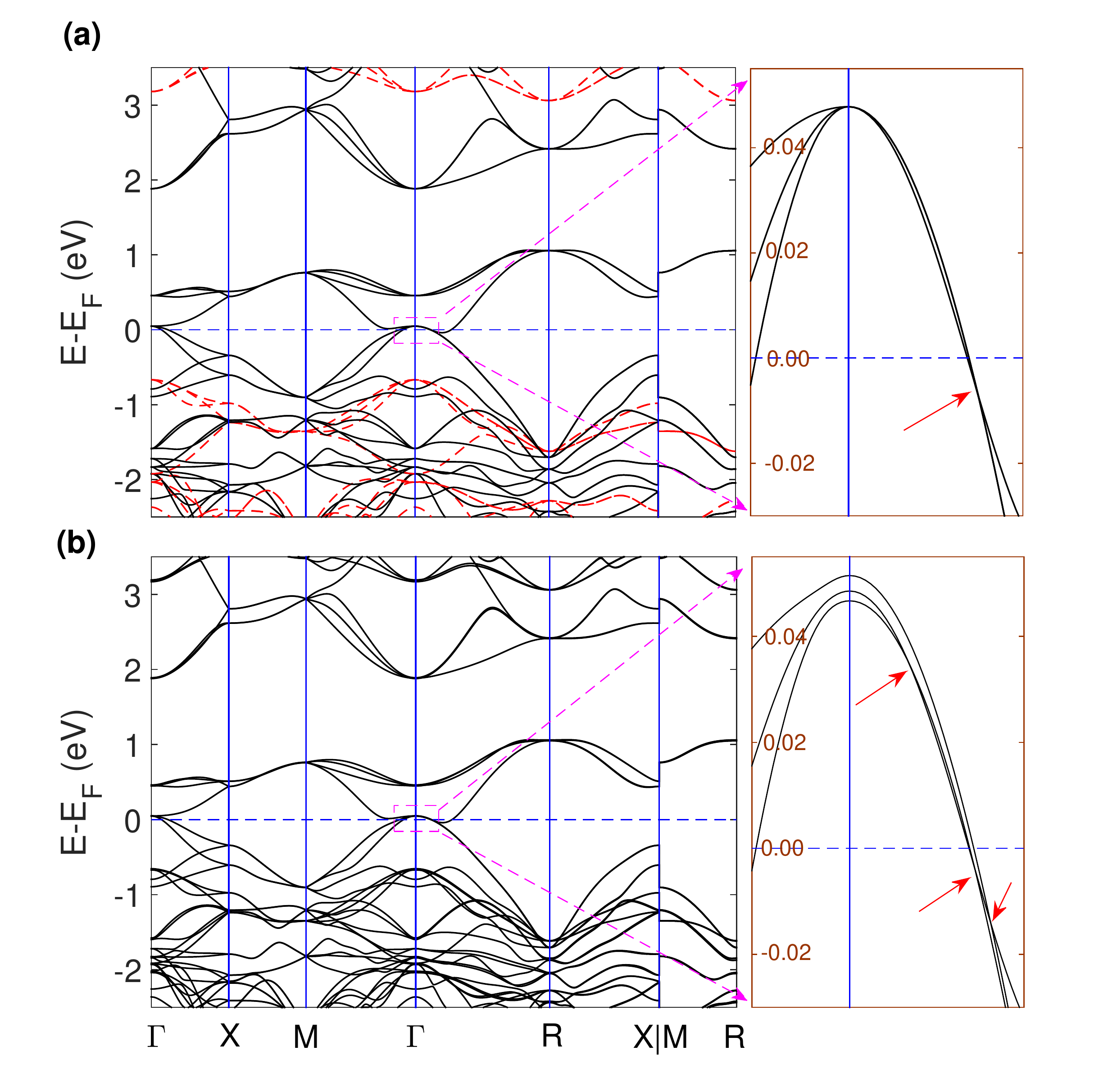}
	\caption{Electronic band structures of pyrite CrO$_2$. (a) The band structures along high-symmetry lines without spin-orbital coupling (SOC). The majority and minority spin bands are denoted by the solid (black) and dashed (red) lines. (b) The SOC band structure  along high-symmetry lines with magnetization along the [111] direction. The enlarged view of the band structures near $\Gamma$ point is shown in the right, and the arrows indicate the corresponding nodal points.
 \label{figure2}}
\end{figure}

\begin{figure}
	\centering
	\includegraphics[scale=0.43]{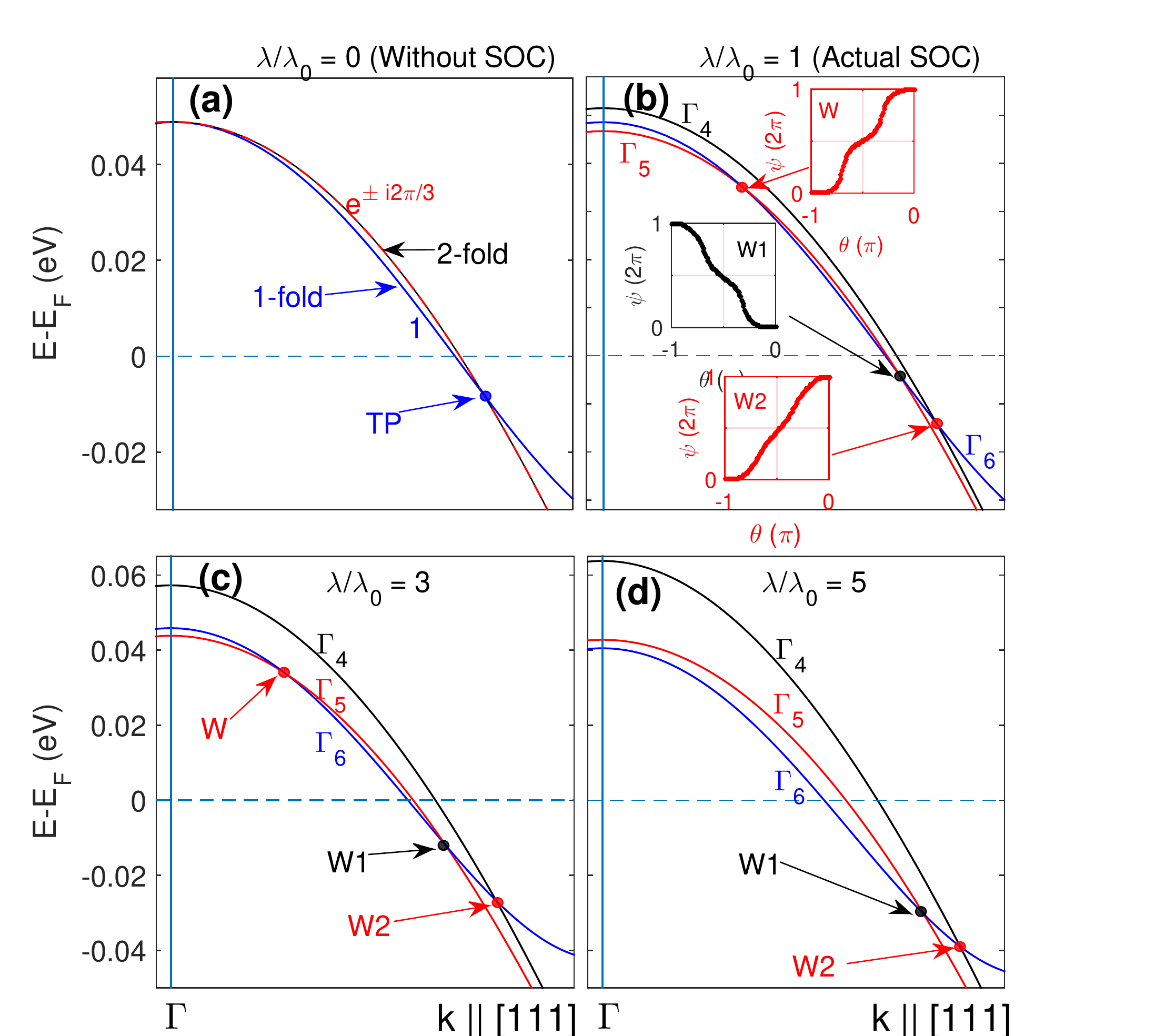}
	\caption{The band topology along $\Gamma$-R direction (or $\mathbf{k}$ $\parallel$ [111]) near $\Gamma$ point. (a) In the absence of SOC, a triply-degenerate node, which originates from the band crossing of a single degenerate band (1) and a two-fold degenerate band ($e^{\pm i \frac{2\pi}{3}}$), is present below Fermi level $\sim$8 meV. (b) In the presence of SOC, the two-fold degenerate band splits into two single-degenerate bands, which then cross the previous single-degenerate band, forming three type-II WPs as $W$, $W1$, and $W2$.  In insets, the Chern number of WP is determined by the evolution of average position of Wannier centers obtained by the Wilson-loop method applied on a sphere that encloses on a WP. (c) and (d) represent tuning the relative SOC strength $\lambda / \lambda_{0}$, where $\lambda_{0}$ is the actual SOC strength.  The increasing of SOC strength would lower the energies of $\Gamma_6$, so the WP $W$ is unstable , while $W1$ and $W2$ are robustly stable on [111] axis.
\label{figure3}}
\end{figure}

As it is illustrated in Fig. \ref{figure1}(a), the pyrite CrO$_2$ crystalizes in simple-cubic (SC) lattice with symmorphic space group $Pa\bar{3}$ (No. 205) that is the semidirect of the point group $T_h$ with the group $\mathbb{Z}^3$ generated by three lattice translations. The optimized lattice constant $a=4.823$ {\AA}, which is good agreement with the previous calculations \cite{Li2012}. The crystal structure consists of interpenetrating Cr and O sublattices, which are located at Wyckoff positions $4a$ (0.0, 0.0, 0.0) and $8c$ (0.3615, 0.3615, 0.3615), respectively. The SC BZ and the projected surface BZ for both (001) and (110) surfaces are shown in Fig. \ref{figure1}(b), in which high-symmetry points are marked.

Our first-principles calculations confirm that the FM state of pyrite CrO$_2$ is considerably more stable that the nonmagnetic and AFM states. The energy of FM state is about 220 meV per Cr atom lower than AFM state, meaning the high Curie temperature above room temperature of this compound. The calculated magnetic moment is $\sim$2 $\mu_B$ per Cr atom, which shows excellent agreement with that in rutile CrO$_2$ \cite{Schwarz1986,Katsnelson2008}. In Figs. \ref{figure2}(a) and (b), the electronic band structures without and with SOC are illustrated, respectively. In the spin-polarized calculations without SOC, the spin and the orbitals are independent.  The minority spin states exhibit the behavior of semiconductor with band gap $\sim$2.9 eV, while the majority spin states show the metallic features, resulting in the half-metallic ferromagnetism of pyrite CrO$_2$. After SOC is present, the calculated band structure in Fig. \ref{figure2}(b) shows the SOC has little influence on the electronic structure and the half-metallic ferromagnetism due to the weak SOC strength of both Cr and O. Furthermore, since the symmetry of FM system is strongly affected by magnetization direction, we perform the first-principles total energy calculations to determine the spontaneous magnetization axis. The [111] easy axis is found to be the energetically most favorable magnetization direction. More interestingly, our calculations show tiny energy differences among all magnetic configurations, implying that an applied magnetic field can easily manipulate the topological states along different directions of magnetization. In the main text, we assume the magnetization is along the [111] direction. The topological analysis of the magnetization along [001] directions is presented in Supplemental Material (SM) \cite{SM}.

As shown in Fig. \ref{figure2}, the topological band crossing occur on the high-symmetry $\Gamma$-R near $\Gamma$ point. In the absence of SOC, the spatial crystal symmetry have no effect on the spin degree of freedom, and two spin channels are decoupled. The symmetry group is the abelian group $T_h$, which contains four three-fold rotational symmetry $C_3$ axes [111], $[1\bar{1}1]$, $[11\bar{1}]$, and $[\bar{1}11]$  plus inversion $I$, and three mirror symmetries $M_x$, $M_{y}$, and $M_z$, respectively. Once SOC is considered, the two spin states couple together and symmetries would decrease depending on the magnetization direction. As the magnetization is along [111] axis, the mirror reflection symmetries $M_x$, $M_{y}$, and $M_z$ are broken. In this case, only the $C_3$ symmetry around [111] axis and $I$ symmetry are still the magnetic symmetry.

We first illuminate the band topology in the absence of SOC. From the magnified band structure along $\Gamma$-R (or $\mathbf{k}$ $\parallel$ [111]) near $\Gamma$ point in Fig. \ref{figure3} (a), we can observe a nodal point below Fermi level $\sim$8 meV is present. This crossing point is a triply-degenerate node, which originates from the band crossing of a single degenerate band and a two-fold degenerate band \cite{Zhuprx2016,Changarxiv}. In this case, three-fold rotational symmetry along [111] axis $C_3^{111}$ and mirror reflection symmetry $M=M_x M_y M_z$ can not commute each other \cite{SM}. Every momentum point on the $\Gamma$-R axis is invariant under the $C_3 ^{111}$ and the product ($IM$) of inversion $I$ and mirror reflection symmetry $M$, meaning that the Bloch states at each point along this direction are also invariant. Therefore, there is always a single degenerate band and a two-fold degenerate band along $\Gamma$-R axis \cite{Changarxiv}. These bands on this axis can be classified by eigenvalues $e^{\pm i \frac{2\pi}{3}}$ and 1 of $C_3$ symmetry. The single degenerate band and two-fold degenerate one can cross each other, and then a pair of triply-degenerate points (TPs) related by $I$ symmetry are present on [111] axis, because the different eigenvalues of $C_3$ symmetry prevent their hybridization. Considering the $T_h$ symmetry, the TPs also occur in $[1\bar{1}1]$, $[11\bar{1}]$, and $[\bar{1}11]$ axes.

\begin{table}
\caption{Nodal points on $\Gamma$-R (or $\mathbf{k}\parallel [111]$) axis of pyrite CrO$_2$. TP and WPs occur in the cases without and with SOC, respectively.
The positions (in reduced coordinates $k_x$, $k_y$, and $k_z$), Chern numbers, and the energies relative to $E_F$ are listed.
$W1$ and $W2$ with opposite Chern numbers are the type-II WPs formed by the splitting of TP in the presence of SOC. The coordinates of the other WPs are related to the ones listed by the inversion symmetry $I$.}
  \begin{tabular}{p{1.0 cm}|*{1}{p{4.0cm}} *{3}{p{1.4cm}} }
  \hline
  \hline
    Nodal  & \centering Coordinates [$k_x(2\pi/a)$,    &\centering  Chern & $E-E_F$ \\
    points & \centering $k_y(2\pi/a)$, $k_z(2\pi/a)$]  &\centering number & (meV) \\
  \hline
    TP  &\centering  (0.0635, 0.0635, 0.0635) &\centering - &{\centering -8} \\
    W  &\centering  (0.0269, 0.0269, 0.0269) &\centering $+1$ & +34 \\
    W1 &\centering (0.0621, 0.0621, 0.0621)  &\centering $-1$ &{\centering -5}  \\
    W2 &\centering (0.0687, 0.0687, 0.0687)  &\centering $+1$ &{\centering -17} \\
  \hline
  \hline
  \end{tabular}
  \label{table}
\end{table}

In the presence of SOC, with magnetization along the [111] direction, the symmetry of pyrite CrO$_2$ is reduced to the magnetic double group $S_6 (-3)$.  The corresponding magnetic space group with [111] magnetization direction contains only six elements formed by two generators: inversion $I$ and $C_3$ symmetry along [111] axis. Since the SOC effect is similar to apply a magnetic field on [111] direction, the magnetization induces that the mirror reflection symmetries $M_x$, $M_{y}$, and $M_z$ are broken.  With magnetization along this axis, the states can be distinguished by the eigenvalues of three-fold symmetry $C_3$ as $e^{i \frac{\pi}{3}}$ for $\Gamma_4$,  $e^{-i \frac{\pi}{3}}$ for $\Gamma_5$, and $e^{-i \pi}$ for $\Gamma_6$, respectively. The effective Zeemann field of SOC leads to the two-fold degenerate band splitting into two single-degenerate bands, each of which corresponds to either of the two irreducible representations $\Gamma_4$ and $\Gamma_5$.  As shown in Fig. \ref{figure3}(b), the band belonging to irreducible representations $\Gamma_6$ would crosses with $\Gamma_4$ and $\Gamma_5$ bands , forming a pair of WPs as $W1$ and $W2$ with opposite Chern numbers. Another WP $W$ induced by the band crossing between $\Gamma_5$ and $\Gamma_6$ is also present.  All WPs located on the $\Gamma$-R axis belong to type-II. The Chern numbers are determined by the evolution of the average positions of Wannier charge centers using \textit{Z2Pack} software \cite{Gresch2017}. The Wilson-loop method applied on a sphere around WPs \cite{Yu2011,Soluyanov2011} is employed as shown in the insets of Fig. \ref{figure3} (b). Their precise positions in momentum space, Chern numbers, and the energies related to the Fermi level $E_F$ are listed in Table \ref{table}. With enhancing the strength of SOC, we can clearly see that the Zeemann splitting increases the distance between $W1$ and $W2$ in momentum space in Figs. \ref{figure3} (c) and (d). Furthermore, the increasing of SOC strength would lower the energies of band $\Gamma_6$, so the WP $W$ is unstable and may be removed, while $W1$ and $W2$ are robustly stable on [111] axis [shown in Figs. \ref{figure3}(d)].

Moreover, pyrite CrO$_2$ exhibits the FM metallic rather than semimetallic features. It is important to note that the nontrivial properties in topological metals not only depend on the relations between valence and conduction bands, since the occupied states are a function of crystal momentum $\mathbf{k}$ in this case \cite{Wang2016prl2}. For instance, the WP $W2$ is the crossing between the $N$th and $(N+1)$th bands, where $N$ is the total number of valence electrons per unit cell of pyrite CrO$_2$. The WPs $W$ and $W1$ are formed by the crossings of the valence bands $(N-1)$ and $N$. In addition, between the bands $(N-1)$ and $N$ we find more additional topologically protected WPs, some of which (6 in total) are close to Fermi level $E_F$. The detail information are supplied in SM \cite{SM}. Importantly, there are only one pair of WPs $W2$ forming at the boundary of electron and hole pockets, while all the other WPs arise at the crossing points of two pockets of the same carriers. Therefore, the surface Fermi arcs from the projections of two type-II $W2$ WPs may be clearly and would be easy to observe in experiments.

\begin{figure}
	\centering
	\includegraphics[scale=0.072]{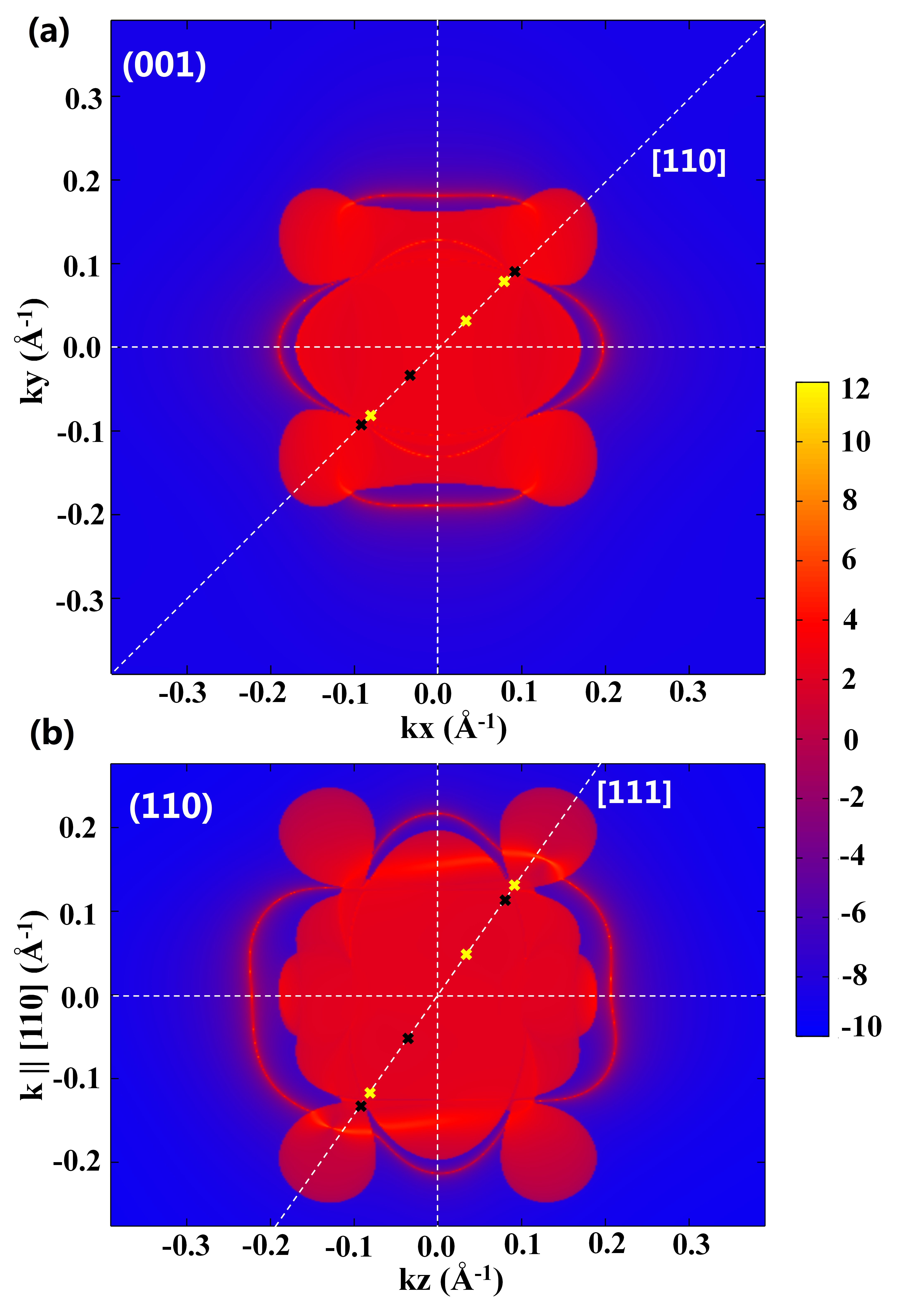}
	\caption{Spectral function of the (a) (001) and (b) (110) surfaces of pyrite CrO$_2$ at 15 meV below the Fermi level $E_F$. We can clearly observe the Fermi arcs arising from WPs forming from the $N$th and $(N+1)$th bands, while the other Fermi arcs are hidden within the projections of bulk pockets. The projections of WPs are denoted by yellow and black crossings, indicating Chern numbers +1 and -1, respectively.
  \label{figure4}}.
\end{figure}

One evident consequence of WPs in pyrite CrO$_2$ is the existence of topologically protected Fermi arcs projected on the surface of this compound. For type-I WPs, the surface Fermi arcs connect the projections of the opposite chirality WPs onto the surface when Fermi level $E_F$ is tuned to the energy of WP. Type-II WPs are located at the boundary the hole and electron pockets, so the open Fermi surface leads to that the projection of WP is always hidden within the projection of bulk pockets on this surface. However, the Fermi arcs can be revealed by tuning the chemical potential \cite{Wang2016prl2}. We consider the (001) and (110) surfaces of pyrite CrO$_2$ to calculate the surface Fermi arcs, since the WPs projected onto these surfaces is distinct in the corresponding surface BZ. The surface density of states of  pyrite CrO$_2$ is computed by using Wannier TB Hamiltonian \cite{Marzari2012,Mostofi2008} with the iterative Green's function method \cite{Sancho1985} as implemented in \textit{Wannier}$_{-}$\textit{tools} package \cite{Wanniertool, Wuwannier}. The Fermi surfaces projected onto (001) and (110) surfaces with the chemical potential 15 meV below Fermi level $E_{F}$ are shown in Figs. \ref{figure4}(a) and (b), respectively. We can clearly observe the Fermi arcs arising from WPs $W2$ of the crossing of $N$th and $(N+1)$th bands, while the other Fermi arcs are hidden within the projections of bulk pockets. The clean Fermi arcs would like to be revealed in angle-resolved photoemission spectroscopy experimentally.

In conclusion, we suggest that pyrite CrO$_2$ co-existing with FM ground state can realize only one pair of type-II WPs between the $N$th and $(N+1)$th bands, where $N$ is the total number of valence electrons per unit cell. Further calculations show that other WPs between the $(N-1)$th and $N$th bands also appear close to the Fermi level due to the complex topological electronic band structure. The symmetry analysis shows that the Weyl points arise from a triply-degenerate point splitting due to the mirror reflection symmetry broken in the presence of SOC, which is equivalent to an applied magnetic field along the direction of magnetization. The corresponding Fermi arcs projected on both (001) and (110) surfaces are calculated as well and observed clearly. Considering that CrO$_2$ is a very common material, this finding opens a wide range of possible experimental realizations of time-reversal breaking type-II Weyl fermions at room temperature.

This work is supported by the National Natural Science
Foundation of China (NSFC, Grant Nos.11204185, 11304403,
11334003 and 11404159).\\
\\
\textbf{Equal Contributions:}\\
R. Wang and Y. J. Jin contributed equally to this work.

\clearpage
\newpage

\setcounter{figure}{0}
\makeatletter

\makeatother
\renewcommand{\thefigure}{S\arabic{figure}}
\renewcommand{\thetable}{S\Roman{table}}
\renewcommand{\theequation}{S\arabic{equation}}

\begin{center}
	\textbf{
		\large{Supplemental Material for}}
	\vspace{0.2cm}
	
	\textbf{
		\large{
			``Ferromagnetic Type-II Weyl Semimetal in Pyrite Chromium Dioxide" }
	}
\end{center}

In this Supplemental Material, we provide the stability analysis of pyrite Chromium Dioxide (CrO$_2$), the other Weyl points between $(N$-$1)$th and $N$th bands, and the topological features with magnetization along [001] direction. Finally, we also elucidate the topology of the triply-degenerate points in the absence of spin-orbital coupling.

\section{The phonon of pyrite chromium dioxide at the ambient pressure}
\begin{figure}
	\centering
	\includegraphics[scale=0.3]{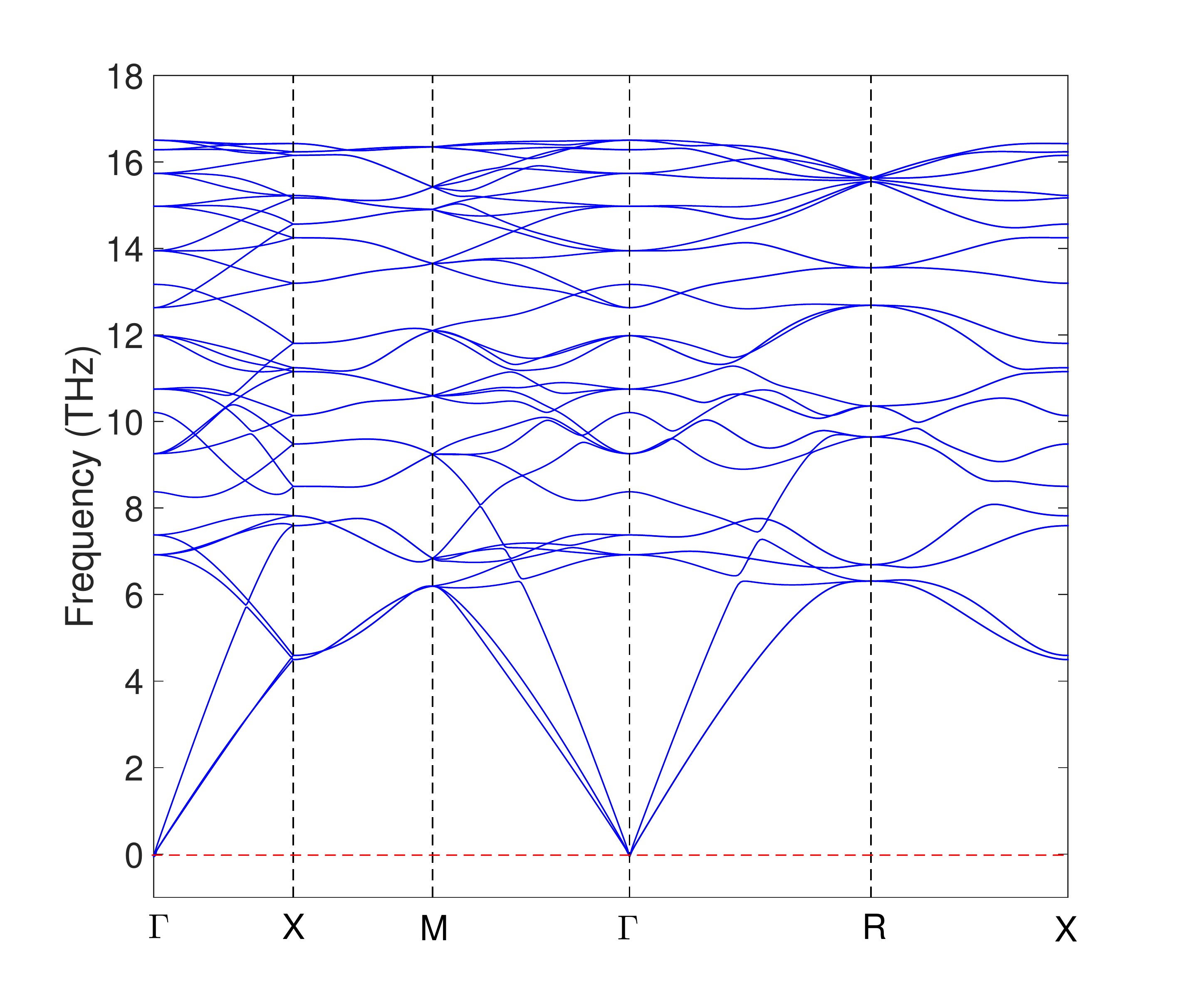}
	\caption{The phonon dispersions of pyrite CrO2.
  \label{figS1}}.
\end{figure}

The pyrite CrO$_2$ phase has been demonstrated to be stable ferromagnetic (FM) half-metallic state occurring at critical pressure of $\sim$45 GPa \cite{Li2012}. Here, we use the phonon spectrum, which is one useful way to investigate the stability and structural rigidity.  The method of force constants has been used to calculate the phonon frequencies as implemented in PHONOPY package~\cite{Togo1,Togo2,Togo3}. We employ
$3 \times3 \times 3$ supercell with 108 Cr atoms and 216 O atoms to obtain the real-space force constants. Our result for the phonon dispersions at the ambient pressure is shown in Fig.\ref{figS1}, respectively. We find that there is the absence of any imaginary frequencies over the entire BZ, demonstrating that the pyrite CrO$_{2}$ is dynamical stability.

\section{The other Weyl points close to Fermi level between $(N$-$1)$th and $N$th bands }
Pyrite CrO$_2$ exhibits the FM metallic rather than semimetallic features. Due to its complex topological electronic band structure, some other Weyl points between the $(N-1)$th and $N$th bands also appear close to the Fermi level. We find that there are three pairs of the Weyl points (the energies relative to Fermi level are lower than 0.3 eV). Their positions in momentum space are $\frac{2\pi}{a}$(0.0011, 0.1414, 0.0842), $\frac{2\pi}{a}$(0.1419, 0.0844, 0.0010), and $\frac{2\pi}{a}$(0.0907, 0.0006, 0.1613), which are of the most relevant located only 0.111, 0.137, and 0.141 eV below the Fermi level, respectively. Although we found a plethora of topological features formed by the $(N-1)$th and $N$th bands, these additional Weyl points and their associated Fermi arcs may overlap with the bulk states when projected onto a surface, such as (001) or (110) surfaces. Hence, these Weyl points can not contribute visible spectroscopic signatures of surface Fermi arcs.

\section{The topological features with magnetization along [001] direction}
Our first-principles calculations suggest that there are only tiny energy differences among all magnetic configurations in pyrite CrO$_2$, implying that an applied magnetic field can easily manipulate the spin-polarized direction. Therefore, we also perform the calculations for magnetization along [001] direction. When the FM magnetization is parallel to [001] direction, the system reduces to magnetic space group $D_{4h}(C_{4h})$ and the three-fold rotational symmetry $C_3$ is broken. Hence, the Weyl points arising from the triply-degenerate points splitting may not locate on $\Gamma$-R axis. In this case, we only pay attention to the Weyl points between $N$th and $(N+1)$th bands. There are five pairs of Weyl points formed at the boundary of electron and hole pockets. Furthermore, the presence of an odd number of pairs of Weyl points between $N$th and $(N+1)$th bands can be clarified by the product of the inversion eigenvalues of the number of occupied bands $N$ at eight time reversal invariant momenta  points $k_{\mathrm{inv}}$ \cite{Hughes2011}, as

\begin{equation}
\chi_{P}=\prod\limits_{k_{\mathrm{inv}};i\in \mathrm{occ}} \zeta_i (k_{\mathrm{inv}}).
\end{equation}

Our calculations show that the value of $\chi_{P}$ is -1, implying that the system may be WSM co-existing with an odd number of pairs of Weyl points. In pyrite CrO$_2$ with magnetization along [001] direction, five pairs of Weyl points between $N$th and $N+1$th bands are present.  Their precise positions in momentum space, Chern numbers, and the energies related to the Fermi level $E_F$ are listed in Table \ref{tableS}.

\begin{table}
\caption{
The Weyl points between $N$th and $(N+1)$th bands with magnetization along [001] direction. The positions (in reduced coordinates $k_x$, $k_y$, and $k_z$), Chern numbers, and the energies relative to $E_F$ are listed.
The coordinates of the other WPs are related to the ones listed by the $I$ symmetry.}
  \begin{tabular}{p{1.0 cm}|*{1}{p{4.0cm}} *{3}{p{1.4cm}} }
  \hline
  \hline
    Weyl  & \centering Coordinates [$k_x(2\pi/a)$,    &\centering  Chern & $E-E_F$ \\
    points & \centering $k_y(2\pi/a)$, $k_z(2\pi/a)$]  &\centering number & (meV) \\
  \hline
    1 &\centering (0.0821,    0.0,   -0.0549) &\centering $-1$  &{\centering 19} \\
    2 &\centering (0.0,    0.0,    0.0159) &\centering $+1$     & 45 \\
    3 &\centering (0.0821, 0.0, 0.0549)     &\centering $+1$       &{\centering 19}  \\
    4 &\centering (0.0,   0.0549,   0.0821)  &\centering $-1$   &{\centering 19} \\
    5 &\centering (0.0    0.0549,   -0.0821)  &\centering $+1$   &{\centering 19} \\
  \hline
  \hline
  \end{tabular}
  \label{tableS}
\end{table}

\section{The triply-degenerate points in the absence of spin-orbital coupling }
In the absence of spin-orbital coupling, the symmetry group is the abelian group $T_h$, which contains four three-fold rotational symmetry $C_3$ axes [111], $[1\bar{1}1]$, $[11\bar{1}]$, and $[\bar{1}11]$, inversion $I$, and three mirror symmetries $M_x$, $M_{y}$, and $M_z$, respectively. The mirror symmetries send
\begin{equation}
\begin{split}
 M_x: (x, y, z) \rightarrow (-x, y, z),\\
 M_y: (x, y, z) \rightarrow (x, -y, z),\\
 M_z: (x, y, z) \rightarrow (x, y, -z),\\
\end{split}
\end{equation}
 and $C_3 ^{111}$ and the product $IM_x M_y M_z$ of inversion $I$ and mirror reflection symmetries leave every momentum point invariant along $\Gamma$-R (or $\mathbf{k}\parallel [111]$) axis. Hence, at each point along the  $\Gamma$-R axis, the Bloch states that form a possibly degenerate eigenspace (band) of the Hamiltonian must be invariant under $C_3 ^{111}$ and $IM_x M_y M_z$. Without SOC, there are three eigenvalues of $C_3$ rotational symmetry, namely, $e^{-i \frac{2\pi}{3}}$, $e^{i \frac{2\pi}{3}}$, and  1 ($e^{i\pi}$), and we denote the corresponding eigenstates as $\psi_{1}$, $\psi_{2}$, and $\psi_{3}$, respectively. Using the basis ($\psi_{1}$, $\psi_{2}$, $\psi_{3}$), the representations of a operators $O$ can be determined as
 \begin{equation}
 O_{ij}=\langle \psi_{i}|O|\psi_{j}\rangle,
 \end{equation}
so $C_3 ^{111}$ and mirror symmetries $M_x$, $M_y$, and $M_z$ can be expressed as
\begin{equation}
C_3 ^{111}=\mathrm{diag}\{e^{-i \frac{2\pi}{3}}, e^{i \frac{2\pi}{3}}, 1\},
\end{equation}
\\
\begin{equation}
M_x=\left(
      \begin{array}{ccc}
        0 & -1 & 0 \\
        -1 & 0 & 0 \\
        0 & 0 & 1 \\
      \end{array}
    \right),
\end{equation}
\begin{equation}
M_y=\left(
      \begin{array}{ccc}
        0 & 1 & 0 \\
        1 & 0 & 0 \\
        0 & 0 & 1 \\
      \end{array}
    \right),
\end{equation}

\begin{equation}
M_z=\left(
      \begin{array}{ccc}
        1 & 0 & 0 \\
        0 & 0 & 1 \\
        0 & 1 & 0 \\
      \end{array}
    \right).
\end{equation}
It can be seen that $C_3 ^{111}$ and $M_i$ ($i=x$, $y$, $z$) can not commute with each other, leading to that failure of $C_3 ^{111}$ and $M_i$ to be simultaneously diagonalizable. Therefore, in the absence of SOC, along $\Gamma$-R (or $\mathbf{k}\parallel [111]$) axis, the three bands with the three different eigenvalues of $C_3 ^{111}$  always appear as a singly-degenerate band ($\psi_{3}$)  and a doubly-degenerate band ($\psi_{1}$ and $\psi_{2}$). If the single degenerate and the
doubly-degenerate bands cross each other accidentally, a triply-degenerate node will form because their different $C_3 ^{111}$  eigenvalues prohibit hybridization \cite{Changarxiv}. When spin-orbital coupling is considered, the triply-degenerate node would like to split into Weyl points depending on the magnetic space group.

\end{document}